# Evaluating the Usability of Qualified Electronic Signatures: Systematized Use Cases and Design Paradigms

Mustafa CAGAL

Informatics Institute, Istanbul Technical University, Istanbul, 34467, Türkiye, cagal21@itu.edu.tr

Kemal BICAKCI

Informatics Institute, Istanbul Technical University, Istanbul, 34467, Türkiye, kemalbicakci@itu.edu.tr

Despite being legally equivalent to handwritten signatures, Qualified Electronic Signatures (QES) have not yet achieved significant market success. QES offer substantial potential for reducing reliance on paper-based contracts, enabling secure digital applications, and standardizing public services. However, there is limited information on their usability despite the extensive range of use cases. To address this gap, we systematize QES use cases and categorize the system designs implemented to support these use cases, emphasizing the necessity to evaluate their respective strengths and weaknesses through usability studies. Additionally, we present findings from cognitive walkthroughs conducted on use cases across four different QES systems. We anticipate that this work will serve as a foundation for a significant expansion of research into the usability of Qualified Electronic Signatures.

## 1 INTRODUCTION

Qualified Electronic Signatures (QES) have been on stage for more than two decades [1]. They are legally equivalent to handwritten signatures and are the mechanism to ensure the authenticity and integrity of electronic data [1, 2, 3]. They have many advantages over handwritten signatures. For instance, they are fast and easy to use, can enhance the pace of business, and reduce the need for paper and pen for organizations. In addition, they provide integrity protection, authenticity, and non-repudiation high-level security goals. They are deployed in various applications by both the public and private sectors today. However, as opposed to their comprehensive use cases, the usability studies are limited.

The global digital signature market size was valued at $3.56 billion and is projected to reach $61.91 billion by 2030, growing at a CAGR (Compound Annual Growth Rate) of 33.2% from 2021 to 2030 [4]. However, despite their vast potential, Qualified Electronic Signatures (QES) have not yet achieved significant market success, as many organizations continue to rely on handwritten signatures due to both usability problems in technology implementation and compliance with laws and regulations. Our primary aim is to point out the usability issues of some QES systems, which may hinder their implementation and the strengths of other QES systems that can be utilized to improve QES to be a more attractive option.

In some countries like the USA and China, while there are electronic signature laws and regulations, the specific implementation of Qualified Electronic Signature (QES) systems is not clearly defined. Conversely, in the EU, the eIDAS Regulation governs QES, outlining available systems that include remote signatures, along with traditional methods like QES tokens and mobile signatures. In contrast, Türkiye's valid QES systems do not incorporate remote signatures. Thus, our research focuses on the QES systems in Türkiye and the EU. We examine the use cases of QES, categorize the system designs for these use cases, and present findings based on cognitive walkthroughs conducted on these use cases.

Our research questions in this paper are as follows:
1. What are the Qualified Electronic Signatures (QES) use cases specifically across the Türkiye and European Union? While using the term "use cases", we refer to the set of tasks that standard users must accomplish.
2. What are the design paradigms, options, sub use cases associated with these use cases?
3. Considering these use cases and design paradigms, what are the strengths and weaknesses as well as the usability challenges of practical QES systems?

The rest of the paper is organized as follows: Our methodology and assumptions are presented in Section 2. Section 3 elaborates on the QES term. Key actors involved in QES processes and related work are presented in Section 4 and Section 5, respectively. Section 6 contains the use cases, sub use cases, and design paradigms. Section 7 includes the results of cognitive walkthroughs. Limitations and future work directions are presented in Section 8, and finally Section 9 provides conclusions.

## 2 METHODOLOGY

While selecting our methodology, we were inspired by the systematization of password managers by Simmons and his colleagues [5]. They categorized the password manager use cases (set of tasks), identified the system designs used to support these use cases (design paradigms), and performed cognitive walkthroughs. We followed their methodology and applied it to Qualified Electronic Signatures (QES).

The initial step of our method is discussing the QES term, key actors and related work. Our second step is identifying use cases and system designs implemented to support these use cases (design paradigms) of QES. The third step is applying cognitive walkthroughs to use cases. The final step is to provide a short discussion of the key findings.

A cognitive walkthrough is a measurement method in which experts perform tasks done by users in real life to detect deficiencies. The tasks are predefined, it is simply a simulation of real life [6]. In more precise terms, the cognitive walkthrough method involves experts attempting to behave like standard users to gain insights into user behavior and detect deficiencies. We selected the cognitive walkthrough method for two reasons: First, cognitive walkthroughs are effective tools for detecting usability issues [5], and second, their task-oriented structure enables researchers to test more systems and use cases. This method allowed us to test 4 practical QES systems, 5 use cases, 14 sub use cases, and 36 design paradigms, which exceeds the solutions and tasks examined in most usability studies. As mentioned in the related work, especially in the QES field, usability studies mostly concentrate on a specific task.

Our research was carried out by two authors: one is an academic researcher with a Ph.D. in information systems, and the other is a Master's student who holds the position of an IT Team Leader in an organization. The IT Team Leader is very familiar with the QES landscape, effectively utilizing QES and incorporating its usage within the company as part of his responsibilities. The research was segmented into three sub-processes: identification of QES use cases, categorization and enumeration of system designs, and exploration of weaknesses, strengths, and usability challenges. Initially, each sub-process was carried out individually, followed by discussions to note any disagreements. These disagreements were then addressed through verbal discussions with both standard QES users and in computer science classes. Each sub process was completed prior to researching the next one.

Table 1: Selected QES Solutions

| Selected Solution | Service Provider | Validity (Türkiye) | Validity (European Union) |
|---|---|---|---|
| QES Token | TURKTRUST Information Communications and Information Security Services Inc. | Valid | Valid |
| QES installed in Turkish national ID | Electronic Information Security Inc. (E-GUVEN) | Valid | Valid |
| Mobile Signature | Vodafone Telecommunications Inc. | Valid | Valid |
| Remote Signature | eSignR/XiTrust Secure Technologies GmbH | Not valid | Valid |

To enhance the clarity of our study, we have developed a user persona named Hasan. Hasan is a dual citizen of Germany and Türkiye. In Germany, he is known as Hans to individuals and colleagues. Hasan resides in Türkiye and holds the position of Finance Team Head at a company in Istanbul. This company operates on a global scale, requiring Hasan to travel frequently. He also has financial expenditure responsibilities. Hasan must sign expenditure documents, which are then forwarded to the bank for company transactions to take place. Hasan has an adequate understanding of technology and he can comprehend fast and utilize new technologies as a means of conducting business. Hasan has two mobile numbers: one from a Mobile Service Operator (MSO) in Türkiye and another from an MSO in Germany. Additionally, he efficiently utilizes the e-government structure. We believe that our research encompasses all possible options for Hasan to utilize Qualified Electronic Signatures.

We assume that Türkiye and the EU present a promising research area to evaluate the strengths and weaknesses of the different QES systems. For the purposes of research, as stated in Table 1, the following QES systems have been selected: Turktrust QES token, E-Guven QES (installed in Turkish national ID), Vodafone mobile signature, and eSignR/XiTrust remote signature. Despite not being recognized as a valid method in Turkish laws, remote signatures are recognized as a valid method in the European Union.

## 3 DECODING THE TERM: "QUALIFIED ELECTRONIC SIGNATURES"

Understanding what a Qualified Electronic Signature is crucial. The term "Electronic Signature" is sometimes employed interchangeably with "Digital Signature" [7]. However, these two terms represent different concepts. "Electronic Signature" is a legal term, but "Digital Signature" is a mathematical and cryptographic tool [8]. The definition of "Digital Signatures" by the European Telecommunications Standards Institute (ETSI) is as follows: "Digital signature is data appended to or a cryptographic transformation of a data unit that allows a recipient of the data unit to prove the source and integrity of the data unit [9]." Another good definition is: "Digital signature is an authentication mechanism that enables the sender of a message to attach a unique code that acts as a signature [10]." However, "Electronic Signatures," according to the European Union Electronic Identification, Authentication, and Trust Services (eIDAS) Regulation Article 3-10, are "data in electronic form which is attached to or logically associated with other data in electronic form and which is used by the signatory to sign [11]." Therefore, any method associated with demonstrating intent, such as appending a name at the conclusion of a document or clicking a button qualifies as an electronic signature. However, this type of electronic signatures is defined as "Simple Electronic Signatures", and not necessarily based on digital signatures. However, "Advanced Electronic Signatures" are based on digital signatures [12].

"Advanced Electronic Signatures" represents a more intricate and robust iteration of electronic signatures. According to eIDAS Regulation Article 26, An "Advanced Electronic Signature" shall meet the following requirements:

1. it is uniquely linked to the signatory;
2. it is capable of identifying the signatory;
3. it is created using electronic signature creation data that the signatory can, with a high level of confidence, use under his sole control and
4. it is linked to the data signed there in such a way that any subsequent change in the data is detectable [11].

Thus, it can be asserted that all the prerequisites for "Advanced Electronic Signatures" are of a technical nature. Furthermore, these technical criteria find fulfilment through the implementation of digital signatures.

"Qualified Electronic Signatures" are "Advanced Electronic Signatures" that are created by a qualified electronic signature creation device and are based on a qualified certificate [11];

1. Signature creation devices can be in many forms like ID cards, USB sticks, SIM cards, smart cards, or remote signature creation devices.
2. Qualified certificates are provided by (public and private) providers that have been granted a qualified status by a national competent authority as indicated in the national 'trusted lists' of the EU Member State [13].

The introduction of the term "qualified" becomes relevant when considering "Qualified Electronic Signatures," imparting a distinct legal significance. In essence, we can state that while "Advanced Electronic Signatures" suffice from a technical standpoint, "Qualified Electronic Signatures" go beyond being deemed "technically and legally" sufficient.

## 4 KEY ACTORS

Key actors involved in QES processes are as seen in Figure 1. We assigned numerical labels to connections in Figure 1 and the corresponding text in this section.

**Users:** Users are individuals who utilize qualified electronic signatures.

**Regulatory bodies:** Regulatory bodies are organizations that establish and enforce the legal framework and standards for electronic signatures. They develop and maintain regulations and standards. They also certify and authorize trust service providers ①. European Union, the French Cybersecurity Agency (ANSSI), the Italian Agency for Digital Italy, and the Turkish Information and Communication Technologies Authority (BTK) are instances of regulatory bodies.

**Trust Service Providers (TSP):** Trust Service Providers (TSP) are organizations that offer electronic signature services. Regulatory bodies authorize TSPs to provide qualified electronic signature services. TSPs manage qualified electronic signatures. Their primary responsibilities include verifying the identity of individuals or entities; issuing, renewing, and revoking certificates ②; maintaining a public key infrastructure (PKI); publishing certificates; keeping the status information of certificates up-to-date; preparing revocation lists, and maintaining an archive of expired and revoked certificates. They also offer libraries, APIs, or browser extensions (integrators) that can be integrated into applications ③. Some TSPs offer platforms to users so that they can utilize them during signature creation ④.

**Platforms:** Platforms refer to mediums through which users can create and verify qualified electronic signatures. Platforms can be web or desktop applications. Also, platforms can be TSP-owned ⑤, institutional TSP-integrated ⑥, or third-party TSP-integrated ⑦. TSP-owned applications are hosted on the platforms that are offered by TSPs. A good example is a remote signature service provider which has a web application, and users accomplish tasks through it. Institutional TSP-integrated applications are the institutional (enterprise) platforms that are integrated with TSPs. Third-party TSP-integrated applications are the online ones where documents are signed by third parties. For instance, users are required to install a TSP library via a DLL to provide signatures in Microsoft Office, which means users are required to accomplish TSP integration. In this case, Microsoft Office is an instance of third-party TSP-integrated desktop applications. The distinction between types (TSP-owned, institutional TSP-integrated, or third-party TSP-integrated) could be significant from the privacy point of view. TSP-integrated applications, either institutional or third-party, must communicate with TSPs via libraries, APIs, or browser extensions ⑧.

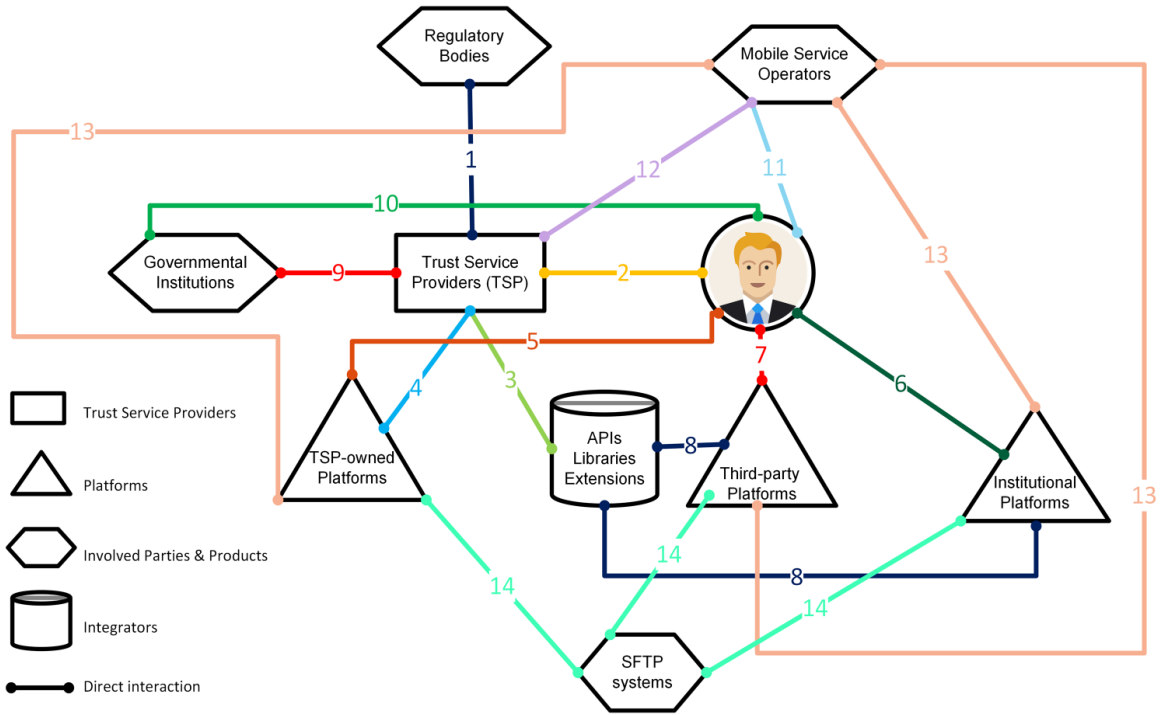

Figure 1: Key actors involved in QES processes.

**Involved parties:** Many other parties (mobile service operators, banks, courts, governmental or educational institutions, public notaries, and e-government systems) are involved with QES processes but do not provide platforms or trust services to users. Noteworthy involved parties are as follows:

QES can be stored in national ID chips. Users apply to trust service providers, TSPs create and manage user certificates and provide service to the devices located in *governmental institutions* ⑨. QES installation requires users to visit governmental institutions ⑩.

QES can also be stored in SIM cards. *Mobile service operators* are not trust service providers, but provide qualified certificates to users. Users apply to mobile service operators; mobile service operators verify users' identity ⑪, send the necessary information to one of the TSPs ⑫, obtain a qualified certificate, and install this certificate on SIM cards. Also, in order to provide mobile signing capabilities, platforms must integrate with mobile service operators ⑬.

Almost every *bank* today has an SFTP (Secure File Transfer Protocol) system that allows users to send QES-signed statements/payment orders. Users create signatures on documents via platforms and then, through platforms send documents to the bank SFTP system ⑭.

## 5 RELATED WORK

Although earlier studies about electronic signatures exist, some focused on technical implementations; others evaluated laws and regulations but few of them analyzed usability issues. Here, we briefly summarize the related work.

Wang [14] examined the e-signature legislative approaches, reviewed the regulations on worldwide and concluded that divergent legislation around the globe can impede the enhancements in the technology. His study consists of four regulations: US, UK, Germany, and China. He stated that all regulations share the same common objective to provide legal validity to entities and facilitate cross-border recognition of them but there is no consensus between different countries on how to achieve this compelling task. He emphasized that international collaboration between national authorities is required to establish a suitable digital environment globally. The assessment of different legislations is beneficial; however, the study only focuses on legal frameworks.

Truong and Minh-Tuan [15] reviewed the usability and other challenges associated with a specific deployment of digital signatures for e-bills in Vietnam. A limitation of the study is that its findings are applicable only to a particular use case (e-bills) in a specific country.

Zefferer and Krnjic [16] made a usability analysis of the Austrian E-government infrastructure based on the electronic signature integration. They compared the card-based qualified electronic signature solutions with mobile signatures. They found that mobile signatures are easier to use from the user's perspective. Similar to Truong and Minh-Tuan's study [14], a limitation of the study is that its findings are applicable only to a particular use case (Austrian E-government infrastructure). Furthermore, their primary focus is not on the electronic signatures but on the aforementioned government infrastructure.

Radka and his colleagues [17] assessed the technical infrastructure of Czech Republic, a European Union member state, and evaluated its readiness. They discussed the consequences of eIDAS from the Czech Republic's perspective. They stated that while electronic signatures are a reality, the commitment, implementation, and employment of QES vary (even) across the EU.

## 6 USE CASES / SUB USE CASES AND DESIGN PARADIGMS

Use cases represent a series of tasks users perform when utilizing Qualified Electronic Signatures. In other words, use cases are primary tasks that users accomplish. Sub use cases reflect a more granular version of these use cases. Finally, design paradigms support these sub use cases. Paradigms represent options for accomplishing the respective sub-use cases. The use cases, sub-use cases, and design paradigms we elaborate on are shown in Table 2 and explained below.

We developed Table 2 considering the infrastructure and services of both Türkiye and the EU. As previously mentioned, QES tokens, mobile signatures, and QES in national ID cards exist both across the EU and in Türkiye. However, our fourth selected solution exists only in the EU. To enhance clarity regarding applicability of our findings, we attach specific tags to the design paradigms:

*TR:* The design paradigm exists only within the Turkish infrastructure (it might be present in the EU but it involves different procedures, making the cognitive walkthrough findings specific to Türkiye).
*EU:* The design paradigm exists only within the EU, making the findings not applicable to Türkiye at this moment.
*TR-EU:* The design paradigm exists within both the EU and Türkiye.

*TR-EU (P):* The design paradigm is relevant to both Türkiye and partially to the EU. The details may vary between Türkiye and the EU and even among EU member states, but the procedures are similar.

Table 2: Use Cases and Design Paradigms (P stands for Partially).

| Use Cases | | Sub Use Cases | | Design Paradigms | TR/EU |
|---|---|---|---|---|---|
| U1 | Obtaining & Setup | U1/1 | obtaining | U1/1-P1 QES token | TR – EU(P) |
| | | | | U1/1-P2 QES in the national ID | TR – EU(P) |
| | | | | U1/1-P3 mobile signature | TR – EU(P) |
| | | | | U1/1-P4 remote signature | EU |
| | | U1/2 | setup computer | U1/2-P1 install programs | TR – EU(P) |
| | | | | U1/2-P2 no installation (online service) | TR – EU |
| U2 | Signature Creation | U2/1 | in web applications | U2/1-P1 TSP-owned applications | TR – EU |
| | | | | U2/1-P2 institutional TSP-integrated applications | TR – EU |
| | | | | U2/1-P3 third-party TSP-integrated applications | TR – EU |
| | | U2/2 | in desktop applications | U2/2-P1 TSP-owned applications | TR – EU(P) |
| | | | | U2/2-P2 institutional TSP-integrated applications | TR – EU(P) |
| | | | | U2/2-P3 third-party TSP-integrated applications | TR – EU |
| U3 | Signature Verification | U3/1 | via TSPs | U3/1-P1 signed with QES token / QES in the national ID / mobile signature | TR |
| | | | | U3/1-P2 signed with remote signature | EU |
| | | U3/2 | via third parties | U3/2-P1 signed with QES token / QES in the national ID / mobile signature | TR |
| | | | | U3/2-P2 signed with remote signature | EU |
| | | U3/3 | via browsers | U3/3-P1 signed in web applications | TR – EU |
| | | | | U3/3-P2 signed in others | TR – EU |
| U4 | Applications | U4/1 | e-government services | U4/1-P1 QES token / QES in the national ID | TR – EU(P) |
| | | | | U4/1-P2 mobile signature | TR – EU(P) |
| | | U4/2 | banking services | U4/2-P1 QES signed orders (SFTP Systems) | TR – EU(P) |
| | | | | U4/2-P2 mobile signature | TR – EU(P) |
| | | U4/3 | email services | U4/3-P1 email signing | TR |
| | | | | U4/3-P2 email verifying | TR |
| | | U4/4 | in other services | U4/3-P1 QES token / QES in the national ID | TR |
| | | | | U4/3-P2 mobile signature | TR |
| U5 | Supplementary Services | U5/1 | lifting block | U5/1-P1 from QES token | TR – EU |
| | | | | U5/1-P2 from QES in the national ID | TR – EU |
| | | | | U5/1-P3 from mobile signature | TR – EU |
| | | | | U5/1-P4 from remote signature | EU |
| | | U5/2 | QES Renewal | U5/2-P1 QES token/ QES in the national ID | TR – EU |
| | | | | U5/2-P2 mobile signature | TR – EU |
| | | | | U5/2-P3 remote signature | EU |
| | | U5/3 | Help Desk & Documentation | U5/3-P1 QES token/ QES in the national ID | TR |
| | | | | U5/3-P2 mobile signature | TR |
| | | | | U5/3-P3 remote signature | EU |

### 6.1 U1 Obtaining & Setup

The first step is obtaining a QES. This use case involves the user's tasks when obtaining the QES and setting up the computer for the first use. Some paradigms necessitate in-person attendance, either at the provider's office or national authority buildings. In contrast, others require a setup phase through mobile service operators. Additionally, some QES types can be obtained entirely online. Also, some QES types require users to download the necessary programs onto their computers, while others can be utilized online.

## 6.2 U2 Signature Creation

This use case contains the user's tasks while creating a signature. Since QES is recognized as equivalent to handwritten signatures by directives and laws, signed documents are turned into legal documents. Some paradigms require users to use web applications, while others require desktop applications.

## 6.3 U3 Signature Verification

This use case involves the user's tasks to verify signed documents. Users can verify via trust service providers, via third parties such as viewers and office suite applications, and in web browsers.

## 6.4 U4 Applications

As the name implies, this use case and respective sub use cases consist the applications that users can utilize QES.

## 6.5 U5 Supplementary Services

This use case contains supplementary services (lifting block, renewal request, help desk services, and documentation) that the user needs while using QES.

# 7 COGNITIVE WALKTHROUGHS

## 7.1 U1 Obtaining & Setup

### 7.1.1 U1/1 obtaining

Qualified Electronic Signatures (QES) can be in the form of USB sticks or cards (USB tokens), or integrated into a SIM (mobile signatures), or linked to national IDs. Lastly, they can be securely stored as remote signatures in the cloud. All the mentioned types are examined as follows:

**(U1/1-P1 QES token)** Qualified Electronic Signatures (QES) stored in USB sticks or cards have been examined as QES tokens in this research due to the similarities in characteristics between these two types of QES. These similarities are:

1. Both can be obtained by users in the same manner.
2. Both require users to carry physical assets, with only a USB stick for the former and a card and reader for the latter.
3. Both of them require the installation of programs.
4. Both may necessitate users to install additional drivers (especially if it involves a sophisticated card reader with keypad and display).
5. Both can be employed in web applications.

The process of obtaining a QES Token involves two sub-processes:

1. Application form fulfillment and checkout
2. Post-payment authentication and contract confirmation

All sub-processes can be physically accomplished in service provider offices. Alternatively, users can complete these two sub processes online.

On provider's websites, an application form exists for users seeking to obtain QES. Users must provide their personal information through the application form, including name and surname, national ID number, mobile phone number, date of birth, email address, address, and invoice details. Upon providing the information, users proceed to the checkout, marking the completion of the first sub-process. All steps in the first sub-process are executed online, and the procedures are straightforward. The second sub-process, namely "Post-payment authentication and contract confirmation," required users to visit either QES offices or public notary until 2021 (in Türkiye). Starting in 2021, users now have three options: (i) visit to a public notary, (ii) a user visit to QES offices or a service provider visit to the users, (iii) online. The options are as follows:

1. The process may involve a public notary. This means the users visit a public notary, get the required documents verifying their identity and provide these documents to the service provider.
2. Users can finalize the process either at QES offices or by requesting the service provider to visit them. When they opt to request the service provider to visit them, a representative from the service provider visits the user. The primary advantage for this case lies in the convenience of acquiring a QES without the need to physically visit somewhere. However, the drawbacks include a higher cost and limited availability of this service, as it may only be offered in some cities.
3. In Türkiye, a new regulation "on the process of verification of the applicant's identity in the electronic communications sector" (2021, Number: 31523) permits providers and users to fulfill the process online [18]. Following the user's completion of the "Application form fulfillment and checkout", the corresponding contract is submitted through the e-government website. However, the sign-in method to the e-government website is crucial. According to regulation, users must sign in using a QES, an internet banking gateway, or a mobile banking gateway to utilize this service [18]. The password authentication method is excluded from the available options due to its security level. Additionally, users seeking a new QES are likely those who still do not have one, making the QES option irrelevant. Therefore, users wishing to utilize this service must sign in to the e-government website with an internet or mobile banking gateway. Upon completing the "post-payment authentication and contract confirmation", users can hand over their QES from designated offices or request the providers to send it by courier.

**(U1/1-P2 QES in the national ID)** The process involves an online pre-application phase. Firstly, users must provide their personal information through the application form and make the payment. Then, users are required to visit the "National General Directorate of Population and Citizenship Affairs Offices" to install QES on their national ID. The installation process requires using "Nüfusmatik" devices located in these offices [19].

The installation process is as follows: Users must first insert their national IDs into "Nüfusmatik" devices and select the QES option. Then, they are required to verify their fingerprint, mobile phone number, and email address. Next, users must approve a letter of commitment that is provided by the service providers. They are then required to enter their national IDs PINs. Ultimately, users are required to generate a PIN for QES.

The directions on "Nüfusmatik" are not well prepared. The installation process can be challenging because users are not adequately informed about the required steps. This may lead to confusion and difficulties during installation, potentially causing frustration.

**(U1/1-P3 mobile signature)** Users must visit authorized mobile service operator offices after completing the pre-application to obtain a mobile signature. Also, to avail of the mobile signatures, users must possess a compatible 128K SIM card. If the SIM card is incompatible, it is exchanged with a mobile signature compatible 128K SIM card [20, 21, 22].

Obtaining mobile signatures can be challenging due to the limited number of authorized offices. For example, one of Türkiye's three major service operators, Vodafone has only 94 authorized offices throughout Türkiye [23]. This means that users may put additional effort to find an authorized office.

**(U1/1-P4 remote signature)** The fully-online process is as follows: The initial step involves completing an online application form that asks for personal details such as name, date of birth, address, email, and phone number. The information entered into the form must match the information on the user's ID or passport. The second step is video identification, which is also completed entirely online with the assistance of a live agent. During this process, the name, address, and date of birth are verified through conversation and a valid ID. Once video identification is completed, the user is required to create a password. This is accomplished by sending an email to the email address specified in the original application form. At the end, a verification code is sent to the mobile phone to finalize the process. However, in our trial, we encountered difficulties since the video identification service provider (Webid, located at Friedrichstraße 88, 10117 Berlin) could not send text messages to a Turkish phone number. As a workaround, a UK mobile phone was used.

*7.1.2 U1/2 setup computer*

Some paradigms require users to download and install the necessary programs onto their computers, while others do not.

**(U1/2-P1 install programs)** Once a user has obtained a QES token, the next step is to download the required programs to their computer. These programs include the Java, .NET framework, and a management program. The Java setup and .NET framework are specific to each 32-bit or 64-bit operating system (OS) [24]. The management program can be installed and launched afterward. This means that users must first determine their OS type. Unfortunately, the documentation and setup videos direct users to the control panel to check their OS and processor type, which can be a burden for users [25]. The host OS can be determined in a way that does not involve users, who are only directed to the required downloads, i.e., online tools such as version checker (https://mdigi.tools/whatversion/) can be used. Also, the multiple-step downloading process can be shortened. Downloads can be conducted automatically with one click user interaction. Additionally, multiple versions of the management programs are available on the provider's websites [24]. The presence of multiple versions poses another challenge for users and results in cognitive overload. Furthermore, the interface design of the management programs seems outdated. Due to the suboptimal interface design, users may experience feelings of insecurity as the interface appears unprofessional. How-to videos and instructions are only for Windows OS versions. Some QES providers state that performing the setup phase on Mac OS is impossible [26, 27]. There are no instructions about other operating systems e.g., Linux distributions.

**(U1/2-P2 no installation (online service))** Some QES types do not require users to download any program. Users who obtain remote signatures can sign documents remotely without the need to install any program and carry additional device. Also, there is no need to set up a computer environment for mobile signatures. However, mobile signatures can be used only with third parties.

**7.2 U2 Signature Creation**

Some use cases require users to utilize web applications, while others require desktop applications.

*7.2.1 U2/1 in web applications*

Web applications can be TSP-owned, institutional TSP-integrated or third-party TSP-integrated. In TSP-owned web applications, users must upload documents to the TSPs' servers whereas third-party TSP-integrated web applications require users to upload documents to third-party servers. These requirements raise privacy concerns. However, in institutional TSP-integrated web applications (institutional document management systems), the documents are signed within the system, and external entities cannot access the documents.

**(U2/1-P1 TSP-owned applications)** Each remote signature service provider has a web application, and users accomplish tasks through it. They are instances of TSP-owned web applications. Users who obtain remote signatures do not need any further registration or setup operations. In our study, the MOXIS platform, which XiTrust Secure Technologies GmbH supports, was used. In this paradigm, the procedure is straightforward. Users need to upload the document and press the sign button. Upon pressing the button, users receive an SMS message consisting of a TAN (Transaction Authentication Number). After entering this TAN into the web application, the document turns into a signed one.

**(U2/1-P2 institutional TSP-integrated applications)** In this paradigm, qualified electronic signatures can be created with QES tokens, QES in the national IDs or mobile signatures.

When users have a QES token or a QES installed national ID, web applications must communicate with USB ports. From web application's perspective, there are two possible ways to establish communication: require users to download a browser extension or require users to download a desktop application. If web application - USB port communication is accomplished via extension, the web application can only serve in specific browser types. Service providers generally use the desktop application option to ensure their web applications are browser-agnostic. After installation of either a desktop application or a browser extension, the web application tries to communicate with the user's certificate. If communication is unsuccessful, the web application requires users to download additional drivers. Once the first communication is established, signing is relatively straightforward. To sign, users press the sign button in the web application and open the desktop application. Users need to connect their tokens or cards. Subsequently, they select the document on the desktop application screen and press OK. Then, the transaction details appear on the screen. The user presses OK to confirm. Finally, they enter their PIN.

Not every web application has signature creation capability with mobile signatures because in order to offer this functionality, service providers must establish integration with mobile signature service operators. However, when integrated, the signature creation procedure requires no setup operation. Users need to input the operator (Turkcell, Vodafone, Turkish Telecom) and their mobile number. Subsequently, they click the sign button. Then, they click OK to confirm the transaction description. Following this, the user observes the fingerprint of this transaction on both their mobile phone and computer screen. The user should verify the fingerprint on their mobile phones and enter their signature PIN.

**(U2/1-P3 third-party TSP-integrated applications)** The procedures are exactly same as institutional TSP-integrated applications, with only one difference. Third-party TSP-integrated web applications require uploading documents to third-party servers, which raise privacy concerns.

*7.2.2 U2/2 in desktop applications*

Desktop applications can be TSP-owned, third-party TSP-integrated, or institutional TSP-integrated.

**(U2/2-P1 TSP-owned applications)** Service providers have developed solutions such as Pinea [28] and Tilia [29] with the capability to create signatures. In some applications, signing with only QES tokens and QES in the national IDs is possible; in others, signing with mobile signatures is also possible.

TSP-owned applications require users to accomplish a setup phase. Users must download the aforementioned TSP-owned desktop application. Java and .NET framework installation are prerequisites. If communication between the application and QES is not successful, users are required to download additional drivers. Then, the procedure of signing varies from program to program but is generally straightforward.

**(U2/2-P2 institutional TSP-integrated applications)** Enterprise desktop applications that support signature creation capabilities are instances of institutional TSP-integrated desktop applications, which utilize APIs to communicate with a TSP. Generally, users are not required to accomplish any setup operation. In some others, users are required to download drivers. The procedure of signing varies from program to program but is generally straightforward.

**(U2/2-P3 third-party TSP-integrated applications)** The instances of third-party TSP-integrated desktop applications include PDF and office suite applications. In our research, both of these types are evaluated.

In PDF applications, QES can be created with only QES tokens or QES in the national IDs. Users with mobile or remote signatures cannot sign via viewers. PDF viewers use the sandboxing technique (i.e., protected view) in default mode to confine the execution environment. So, creating and executing files and modifying system information such as certain registry settings and other control panel functions are prevented [30]. Users must first disable this protected view. Then, users are required to install the Tubitak Akis PKCS#11 module via a DLL. PKCS#11 modules are platform-independent APIs that allow communication between desktop applications and cryptographic tokens [31]. Lastly, users are required to plug in their token or national ID and log in with their PIN. Users select Digitally Sign" and press the Sign button. After pressing the button, the details appear on the computer screen. The users enter the PIN and press OK. Website instructions and how-to videos are only about Adobe Reader. There is a lack of documentation and resources available for users who wish to sign documents but do not use Adobe Reader.

Office documents can be signed with QES tokens or QES in the national IDs. The procedure can be carried out within the Microsoft Office suite. Users must open their document, navigate to "File" - "Info," and choose "Protect Document." Subsequently, users must click on the "Add a Digital Signature" option. Afterwards, users must select a "Commitment Type" and click "Sign." The next step involves users plugging in their token or card, entering their PIN, and pressing OK. Website instructions and how-to videos consist only of legacy Office versions. A how-to video published on a service provider web site specifies that document signing is supported only in Office 2007 and 2010 versions [32]. There is a lack of documentation for users who wish to sign documents on Microsoft 365.

### 7.3 U3 Signature Verification

Our examination specifically focuses on signature verification in this use case.

*7.3.1 U3/1 via TSPs*

TSPs offer web and desktop platforms, which provide verification services. Successful verification relies on the inclusion of the service provider certificate in the trusted certificate database of the validation servers. The document is verified successfully if the QES is listed in the trusted certificate database.

**(U3/1-P1 signed with QES token / QES in the national ID / mobile signature)** QES service operators are listed only on trusted lists of local solutions.

**(U3/1-P2 signed with remote signature)** The remote signature service providers are included both in the TSP certificate databases and global lists such as European Union Trusted Lists (EUTL).

*7.3.2 U3/2 via third-parties*

The instances of third parties include external websites, code repositories, web applications, or desktop applications such as PDF viewers and office suite applications.

A verification service in the Turkish e-government website is limited to verifying selected QES providers (Similarly, Austria's supervisory body for QES has established a central verification website [33]). These services operate on a local level. When an electronically signed document is shared with others and opened with Microsoft Office applications, an automatic signature verification process starts. Microsoft only considers trust service providers included in the Microsoft Trust Center. Similarly, with PDF viewers, an automatic signature verification process starts. The document is verified successfully if a QES service provider is listed in viewer root lists, i.e., Adobe Approved Trust List (AATL) and the European Union Trusted Lists (EUTL) [34] (Similarly, Foxit trust service providers are included in two root lists [35]). We also note that the EU has developed an open-source library named Digital Signature Services [36], released under the Lesser General Public License. The library has the capability to verify certificates listed in the EUTL.

**(U3/2-P1 signed with QES token / QES in the national ID / mobile signature)** QES service providers are listed only on trusted lists of local solutions and none of QES Service Providers are listed in the global lists, trusted lists of viewers, Microsoft, or other third parties. No matter which method, verification fails. Adding certificates to the trusted lists is possible. However, this requires users to add certificates manually; in this situation, users may experience feelings of insecurity.

**(U3/2-P2 signed with remote signature)** The remote signature service providers are included in the global lists (EUTL, AATL, FATL), trusted lists of viewers, Microsoft Trust Center and other third parties.

*7.3.3 U3/3 via browsers*

**(U3/1-P1 signed in web applications)** In most cases, a visual signature mark is embedded into the documents after signing. By default, this mark includes the name, date, and time information about the QES. However, nothing changes visually after signing documents in some applications. So, users cannot differentiate a signed document from a non-signed one.

**(U3/2-P2 signed in others)** When an electronically signed document is opened with browsers, users can see the visual representation of the signature, so it provides information about the signature. This refers to the lowest verification level.

**7.4 U4 Applications**

*7.4.1 U4/1 e-government Services*

If users authenticate themselves on the e-government with QES, they can access additional services. These services include contract confirmation, mobile service operator change application, and registered email address (KEP) application.

**(U4/1-P1 QES token / QES in the national ID)** To accomplish this task, users must first install an application with Java Network Launching Protocol (JNLP). After installation, the token or card must be plugged. Users must enter their identity number into the e-government website to generate an operation code. Then, users must enter this operation code into the JNLP application together with their PIN.

**(U4/1-P2 mobile signature)** Users are required to provide their identity number, mobile phone number, and mobile service operator name. After the user provides inputs and clicks on the sign button, a message appears on the browser indicating that the mobile signature request has been successfully processed. A warning appears on the user's phone screen, displaying the transaction's fingerprint that should match the one on the web page. The users should complete the transaction by entering their PIN.

*7.4.2 U4/2 banking services*

In certain cases, authentication or authorization with QES is possible.

**(U4/2-P1 QES signed orders)** This design paradigm is crucial for the companies. Typically, companies have finance departments responsible for processing payment transactions. To complete transactions, authorized persons must sign the necessary documents, which are then sent to banks. Document delivery to the bank is generally done through the employees of that company or couriers if handwritten signatures are in use. On the other hand, QES-signed order also can be delivered through SFTP systems.

Almost every bank today has an SFTP (Secure File Transfer Protocol) system that allows companies to send QES-signed statements/payment orders. To utilize this service, the requirements are;

1. Users must have a QES token, QES in the national ID or mobile signature.
2. Users must physically visit bank office and register their certificates to the banking system.
3. Users must sign documents on applicable platforms. Not all platforms have the capability to transfer signed documents to the banking system.

**(U4/2-P2 mobile signature)** Some banks allow their users to log in to mobile banking systems with mobile signatures. However, this option is not accessible by default and requires users to accomplish first a setup phase. The setup phase can be accomplished online.

*7.4.3 U4/3 email services*

While QES can technically be utilized to sign emails, it does not usually confer legal status to the signed emails. In accordance with the Electronic Signature Law 5070 enforced in Türkiye, the law requires that secure electronic signature creation tools allow the signer to view all data before creating the signature (Electronic Signature Law 5070-Article 6) [2]. However, typical email service providers include additional data that is not visually accessible to users, such as rich text formatting information, background color codes, HTML information. Therefore, signatures created on emails using popular email service providers like Gmail, Hotmail, Yahoo, and others may not be considered valid under law. To ensure that electronically signed emails are considered valid under the law and deemed valid certificate usage, obtaining a registered email address (the acronym KEP is used in Türkiye) is necessary.

**(U4/3-P1 email signing)** To ensure that electronically signed emails are both considered valid under the law and are deemed as a valid certificate usage method, it is necessary to follow specific requirements. These requirements include:

1. Using a Qualified Electronic Signature (QES)
2. Utilizing registered email address (KEP)
3. Employing the web application provided by a registered email address service provider to compose the emails (A web application is sufficient for receiving and verifying signed emails, but to send signed emails, additionally, the use of a JNLP (Java Network Launch Protocol) application is required.)
4. Ensuring that the recipient also has a registered email address.

The process of obtaining a registered email address differs based on whether the user possesses a QES or not. If a user already has a valid QES, the entire process can be completed online. However, for users who do not have a QES and only wish to receive signed emails, the obtaining process requires an in-person setup. To send signed emails, users need to install an application that works in Java Network Launching Protocol (JNLP). This JNLP application enables an application to be launched on a client desktop by using resources that are hosted on a remote web server [37].

Subsequently, users create emails on the registered email service provider's web application and click sign; then an operation code is created. Users must enter this operation code into the JNLP application and their PIN, respectively. If user inputs are valid, signed emails are sent.

**(U4/3-P2 email verifying)** Utilizing a registered email address and employing the web application linked to the service provider is sufficient to verify signed emails [38]. When an email is sent, what is known as email evidence is created automatically. Email evidence consists of information: when, by whom, which user sent the email; by which user email is accepted; to whose mailbox and when email was delivered; when it was opened and read by whom. Email evidences can be verified and downloaded [38].

*7.4.4 U4/4 other services*

Some government departments in Türkiye provide authentication services to users who have QES. This list includes but is not limited to the Ministry of Justice (National Judicial Network Project - UYAP), Ministry of the Interior, Ministry of Trade, and Turkish Revenue Administration (GİB).

**(U4/4-P1 QES token / QES in the national ID)** and **(U4/4-P2 mobile signature)** The procedures to utilize these services are mostly same as the e-government use case.

**7.5   U5 Supplementary Services**

*7.5.1 U5/1 lifting block*

**(U5/1-P1 from QES token)** QES is blocked when users enter PIN code incorrectly three times. Lifting blocks can be performed via management programs. The procedure is as follows: users are first requested to contact the call center and acquire a verification code. Subsequently, users enter the verification code into the QES management program.
**(U5/1-P2 from QES in the national ID)** The procedure is mostly same as the QES Token paradigm.
**(U5/1-P3 from mobile signature)** Lifting blocks after three incorrect entries can be performed via customer service calls. This procedure may present difficulties due to its infrequent usage, as agents at call centers may not be well-versed.
**(U5/1-P4 from remote signature)** QES is blocked when users enter the signature password incorrectly ten times. Remote signing service providers explicitly state that the signature password cannot be reset for security reasons. In these situations, users must apply again for a new remote signature [39].

*7.5.2 U5/2 QES renewal*

Qualified Electronic Signatures are valid for only a limited lifetime. At the end of the validity period, if the users wish to continue to utilize QES, they need to renew their certificates. Validity periods of one, two, or three years are available.
**(U5/2-P1 QES token / QES in the national ID)** QES Providers inform the users via email when the expiration date is about to happen. Renewal requests can be conducted via management programs or provider websites. The process can be completed online and does not necessitate re-authentication.
**(U5/2-P2 mobile signature)** Mobile signatures have only a one-year validity period. It is necessary to renew it shortly before it expires. Failure to renew the certificate results in revocation of the certificate. An informative SMS is sent 30 days before the certificate expiration date, and users who approve the SMS have their certificates renewed. When the user does not make a renewal request, reminder SMS messages are sent again 20 days and 10 days before the certificate expires.
**(U5/2-P3 remote signature)** Users are informed via email when the expiration date is about to happen. The renewal process starts with the user's application on the provider's website. The process can be completed online and does not necessitate re-authentication.

*7.5.3 U5/3 help desk & documentation*

Documentation, how-to videos, and help desk access are crucial for most users who may not have extensive knowledge of QES. Therefore, the availability and quality of the documentation and how-to videos are essential.

**(U5/3-P1 QES token / QES in the national ID)** The documentation consists of website instructions and how-to videos, but they are not very well-prepared or up-to-date. Service providers operate call centers that are available only on business days from 09:00 to 18:00.

**(U5/3-P2 mobile signature)** The documentation includes only website information and a frequently asked questions (FAQ) section, but no how-to videos are available. Service operators operate call centers that is available 7/24. However, in our experience, we observe that call center agents may need to be better equipped to respond to users' queries and issues.

**(U5/3-P3 remote signature)** The documentation includes mostly up-to-date information and comprehensive how-to videos. The service providers operate call centers available only on business days from 09:00 to 17:00.

## 8 LIMITATIONS AND FUTURE WORK

Due to the differences in the policies, laws, electronic signature levels, and available implementations globally, we focused on the EU and Türkiye in our research. This geographic scope may be considered as a limitation.

Utilizing signature services in alternative operating systems to Windows can be problematic, particularly when qualified electronic signatures are in the form of tokens or cards because the applications must communicate the USB ports using some of the operating system attributes. We conducted our research specifically on Windows, which could be considered as a limitation.

An interesting research path would be to analyze QES systems in other countries, assess their strengths and limitations, and determine their applicability in the EU and Türkiye. Another promising research direction could be conducting user studies to further enhance our understanding of various design paradigms.

## 9 CONCLUSION

Although QES has significant potential, it has yet to achieve substantial success because of usability challenges in its technology implementation. To address this gap, we conducted a comprehensive investigation to identify the usability issues of QES systems, which may hinder adoption. We selected the European Union and Türkiye as the focal points of our research due to the promising potential for evaluating the strengths and weaknesses of different QES systems. We elaborated on the key actors involved in QES processes and discovered that numerous actors in QES processes can affect usability. Then, we identified and categorized the use cases, sub use cases and system designs implemented to support these use cases. Finally, we conducted 36 cognitive walkthroughs in total. Our findings are summarized as follows:

1. QES tokens, QES in the national IDs, and mobile signatures are legally recognized in Türkiye. However, these solutions have more usability limitations compared to remote signatures. Remote signature is legitimate option within the European Union.
2. Both QES tokens and QES in the national IDs necessitate users to carry physical assets, such as USB sticks or cards, card readers. However, remote signatures do not necessitate users to carry physical assets.
3. It is impossible to obtain QES either through national IDs or as mobile signatures via online means. QES in the national ID requires a visit to the National General Directorate of Population and Citizenship Affairs offices, and obtaining mobile signatures requires a visit to authorized mobile service operator offices. Furthermore, there are a limited number of authorized mobile service operator offices.

4. QES tokens can be obtained online but requires a setup. Some of the required programs are specific to each operating system type. Also, device or card reader drivers need to be installed. This requirement prevents the solution from being device-agnostic. However, remote signatures are device agnostic; users can utilize the solution via various devices.
5. QES tokens and QES in the national IDs can be used to create signatures on the web. However, even in web use cases, users must download a browser extension or another desktop application because web applications must communicate with USB ports. Remote signatures are browser-agnostic; users can utilize the solution via various browsers. Users are not required to accomplish additional setup or installation. Extensions, drivers, and desktop applications are not required.
6. Mobile signatures are not independent solutions. They can only be utilized with third parties. However, each remote signing service provider offers its web application, making remote signatures an autonomous option. Third parties are unnecessary.
7. Remote signatures also have an advantage over the others with respect to verification. Service providers are listed in Microsoft Trust Center and European Union Trusted Lists (EUTL), so shared signed documents are automatically verified. However, for others, verification is only possible in local solutions. Automatic signature verifications in third parties fail. This situation may result in feelings of insecurity.
8. The remote signature documentation includes mostly up-to-date information and comprehensive how-to videos. However, the documentation is not well-prepared and up-to-date in other options. Furthermore, for mobile signatures, the available information is limited to website blog content and a frequently asked questions (FAQ) section.
9. QES token and QES in the national ID management programs seem unpolished whereas there is no need to utilize management software for remote signature solutions.
10. Mobile signature requires users to have compatible 128K SIMs. If the SIM card is incompatible, it needs to be exchanged with a Mobile Signature Compatible 128K SIM card.
11. Because remote signatures are not valid in Türkiye at the moment, it is impossible to use them for authentication in E-government or mobile banking.

Our main finding is that remote signatures are more usable compared to other alternatives. Based on our research, we suggest legalizing remote signatures in Türkiye and other countries where they have not yet been authorized, to make qualified electronic signatures a more attractive option.

During the preparation of this work the authors used ChatGPT in order to language improvement. After using this tool, the authors reviewed and edited the content as needed and takes full responsibility for the content of the publication.